\documentclass[a4paper,12pt]{article}
\usepackage{amssymb}
\usepackage{amsmath}
\usepackage{amsthm}
\usepackage{latexsym}
\usepackage{lscape}
\newtheorem{thm}{Theorem}

\begin{document}
\title{\Large\textbf{Automorphisms of Real 4 Dimensional Lie Algebras and the
Invariant Characterization of Homogeneous 4-Spaces}}
\author{\textbf{T. Christodoulakis}\thanks{tchris@cc.uoa.gr} ~~~\textbf{G.O. Papadopoulos}\thanks{gpapado@cc.uoa.gr}\\ University of Athens, Physics Department\\
Nuclear \& Particle Physics Section\\
Panepistimioupolis, Ilisia GR 157--71, Athens, Hellas \and\\
\textbf{A. Dimakis}\thanks{dimakis@aegean.gr}\\ University of the Aegean\\
Dept. of Financial and Management Engeneering\\ 31, Fostini Str.
82100, Chios, Hellas }
\date{}
\maketitle
\numberwithin{equation}{section}
\begin{abstract}
The automorphisms of all 4-dimensional, real Lie Algebras are
presented in a comprehensive way. Their action on the space of
$4\times 4$, real, symmetric and positive definite, matrices,
defines equivalence classes which are used for the invariant
characterization of the 4-dimensional homogeneous spaces which
possess an invariant basis.
\end{abstract}
\newpage
\section{Introduction}
Automorphisms of 3-dimensional, real Lie Algebras \cite{Harvey},
have been proven a powerful tool for analyzing the dynamics of 3+1
Bianchi Cosmological Models \cite{Henneaux}. At the classical
level, time-dependent automorphisms inducing diffeomorphisms can
be used to simplify the line element -- and thus the Einstein's
Field Equations -- without loss of generality \cite{ChrisJMP}.
They also provide an algorithm for counting the number of
essential constants; the results obtained agree for all Bianchi
Types with the preexisting results \cite{ellis} but, unlike these,
the algorithm can be extended to 4 or more dimensions. At the
quantum level, outer automorphisms provide integrals of motion of
the classical Hamiltonian dynamics; their quantum analogues can be
used to reconcile quantum Hamiltonian dynamics with the kinematics
of homogeneous 3-spaces \cite{ChrisCMP}.

A corresponding analysis of these issues for the case of 4+1
spatially homogeneous geometries, seems very interesting in
itself. It could also prove valuable for the nowadays fashionable
brane world models. As a first step in implementing such an
analysis, we exhibit the automorphisms for all, real,
4-dimensional Lie Algebras (a first treatise of the subject can be
found in \cite{Fee}) and subsequently use them, to invariantly
describe homogeneous 4-spaces.

\section{Automorphisms}
Before exhibiting the results on the automorphisms and their
generators, we briefly recall some basic elements of the Theory of
Lie Groups. Topological issues will not concern us, since at this
stage of study, are rather irrelevant.

Let $V_{N}$, be a vector space over the field $\mathbb{R}$. For
each point $x^{m}$ in the space, a set of transformations
$\widetilde{x}^{\,m}=f^{m}(x^{n};\alpha^{\mu})$ depending on some
parameters $\alpha^{\mu}$ (with Greek indices ranging on the
closed interval $[1,\ldots,M]$, while Latin ones, on the closed
$[1,\ldots,N]$) is defined, endowed with the following properties:
\begin{itemize}
\item The parameters are essential, i.e. they are not functions of
others; rather, they take values on a compact domain.

\item There are particular values for each and every parameter
$\alpha^{\mu}$ -- which without loss of generality can be taken to
be all zero -- such that $x^{m}=f^{m}(x^{n};0,\ldots,0)$. In other
words, the identity transformation, is reached continuously, when
all the parameters reach this particular set of values (here the
zeros).

\item The Jacobian of the transformation, $J^{m}_{n}=|\partial
f^{m}(x^{k};\alpha^{\mu})/\partial x^{n}|$, is non vanishing on
its entire domain of definition. For, every transformation, at
least locally, must be invertible.
\end{itemize}

If one assumes that the parameters are small, and expands in
Taylor series the transformations (the functions $f^{m}$ are taken
to be $C^{n}$ differentiable, with $n$ depending on the
application) he will get:
\begin{eqnarray}
\widetilde{x}^{\,m}=f^{m}(x^{k};0,\ldots,0)+\alpha^{\mu}
\left(\frac{\partial f^{m}(x^{k};\alpha^{\nu})}{\partial
\alpha^{\mu}}\Big|_{\alpha^{\nu}=0}\right)+\mathcal{O}(\alpha^{\mu}\alpha^{\nu})
\end{eqnarray}
(the Einstein Summation Convention, is in use). Then, a set of
$M$, $N$-dimensional, vector fields (each for every essential
parameter), is associated to the previous infinitesimal
transformations:
\begin{eqnarray}
X^{m}_{\mu}=\frac{\partial
f^{m}(x^{n};\alpha^{\nu})}{\partial\alpha^{\mu}}\Big|_{\alpha^{\nu}=0}
\end{eqnarray}
These vector fields are called ``Generators'' and form an Algebra:
\begin{eqnarray} \label{algebra}
[X_{\mu},X_{\nu}]=C^{\kappa}_{\mu\nu}X_{\kappa}
\end{eqnarray}
which is called Lie Algebra -- due to the above mentioned
properties. If the quantities $C^{\kappa}_{\mu\nu}$ do not depend
on the space point $x^{m}$, they are called ``Structure
Constants'', otherwise ``Structure Functions'' and the
corresponding algebras, open Lie Algebras. In what follows, the
Lie Algebras, are assumed to be closed. In this case, the vector
space $V_{N}$, admits a Group of Motions (i.e. transformations)
$G_{M}$ to which the Lie Algebra of the generators of the
transformations, is associated. Then it can be proven that
$M<N(N+1)/2$ -- see \cite{Petrov}, for a detailed analysis.

The Jacobi Identities for the generators, hold:
\begin{eqnarray}
[[X_{\mu},X_{\nu}],X_{\kappa}]+[[X_{\nu},X_{\kappa}],X_{\mu}]
+[[X_{\kappa},X_{\mu}],X_{\nu}]=0
\end{eqnarray}
or in terms of the structure constants:
\begin{eqnarray}
C^{\rho}_{\mu\nu}C^{\sigma}_{\rho\kappa}+C^{\rho}_{\nu\kappa}C^{\sigma}_{\rho\mu}
+C^{\rho}_{\kappa\mu}C^{\sigma}_{\rho\nu}=0
\end{eqnarray}
If one contracts the index $\sigma$ with a contravariant index --
e.g. $\kappa$ --, one gets the Contracted Jacobi Identities:
\begin{eqnarray}
C^{\rho}_{\mu\nu}C^{\sigma}_{\rho\sigma}=0
\end{eqnarray}
and thus a ``natural'' quantity emerges, namely
$C^{\sigma}_{\rho\sigma}\doteq \nu_{\rho}$, which is a covector
under the action of $GL(M,\mathbb{R})$ -- see \cite{MacCallum} and
the references therein.

Under a linear mixing, i.e. the action of the $GL(M,\mathbb{R})$,
of the generators:
\begin{eqnarray} \label{transformation}
X_{\nu}\rightarrow \widetilde{X}_{\nu}=L^{\mu}_{\nu}X_{\mu}
\end{eqnarray}
the structure constants, transform -- according to (\ref{algebra})
-- as:
\begin{eqnarray}
C^{\kappa}_{\mu\nu}\rightarrow
\widetilde{C}^{\kappa}_{\mu\nu}=L^{\alpha}_{\mu}L^{\beta}_{\nu}(L^{-1})^{\kappa}_{\rho}C^{\rho}_{\alpha\beta}
\end{eqnarray}
while:
\begin{eqnarray}
\nu_{\nu}\rightarrow \widetilde{\nu}_{\nu}=L^{\mu}_{\nu}\nu_{\mu}
\end{eqnarray}
The subset of those transformations (i.e. of the form
(\ref{transformation})) with respect to which the structure
constants are invariant, is the Automorphism Group of the Lie
Algebra $Aut(G)$. If $\Lambda^{\mu}_{\nu}$ are the matrices of
this group, then:
\begin{eqnarray} \label{automorphism}
C^{\kappa}_{\mu\nu}=\Lambda^{\alpha}_{\mu}\Lambda^{\beta}_{\nu}(\Lambda^{-1})^{\kappa}_{\rho}C^{\rho}_{\alpha\beta}
\end{eqnarray}

At first sight, in order to find the Automorphism Group of a given
Lie Algebra (i.e. for a given set of non vanishing structure
constants), one has to solve the cubic system
(\ref{automorphism}), which can be transformed to quadratic by
noting that the matrices of interest, are non singular and thus:
\begin{eqnarray} \label{automorphism2}
C^{\kappa}_{\mu\nu}\Lambda^{\rho}_{\kappa}=\Lambda^{\alpha}_{\mu}
\Lambda^{\beta}_{\nu}C^{\rho}_{\alpha\beta}
\end{eqnarray}
but still, the quest for the solutions, remains a difficult task.

A first simplification may by achieved by observing that:
\begin{eqnarray} \label{automorphism_vector}
\nu_{\nu}=\Lambda^{\mu}_{\nu}\nu_{\mu}
\end{eqnarray}

A second simplification, makes use of the Killing-Cartan metric --
provided by a famous theorem due to Cartan:
\begin{eqnarray}
g_{\mu\nu}=C^{\alpha}_{\beta\mu}C^{\beta}_{\alpha\nu}
\end{eqnarray}
Automorphisms preserve its form i.e. are isometries of this
metric:
\begin{eqnarray} \label{automorphism_cartan}
g_{\mu\nu}=\Lambda^{\alpha}_{\mu}\Lambda^{\beta}_{\nu}g_{\alpha\beta}
\end{eqnarray}
The combined use of (\ref{automorphism2}),
(\ref{automorphism_vector}) and (\ref{automorphism_cartan}) makes
the first simpler to solve, providing us with useful necessary
conditions restricting the $\Lambda^{\mu}_{\nu}$'s.

In order to find the generators of the Automorphism Group of a
given Lie Algebra, it is necessary to consider a family of
automorphic matrices which is connected to the identity i.e.
$\Lambda^{\mu}_{\nu}=\Lambda^{\mu}_{\nu}(\tau)$ for some parameter
$\tau$ such that $\Lambda^{\mu}_{\nu}(0)=I_{M}$, with $I_{M}$ the
$M$ dimensional identity matrix. Then if one substitutes this
family to (\ref{automorphism2}), differentiates with respect to
this parameter, and sets at zero, one will get:
\begin{eqnarray} \label{generators}
\lambda^{\kappa}_{\rho}C^{\rho}_{\mu\nu}=
\lambda^{\rho}_{\mu}C^{\kappa}_{\rho\nu}+\lambda^{\rho}_{\nu}C^{\kappa}_{\mu\rho}
\end{eqnarray}
where:
\begin{eqnarray}
\lambda^{\mu}_{\nu}=\frac{d\Lambda^{\mu}_{\nu}(\tau)}{d\tau}\Big|_{\tau=0}
\end{eqnarray}
is the requested generator. The system (\ref{generators}) is
linear and thus easy to solve, when the values of the structure
constants are given. The number of independent solutions to it,
determines the number of the independent parameters of the
generators of the Automorphism Group.

The situation in the literature, concerning the 4-dim, real Lie
Algebras, is characterized by a certain degree of diversity. The
main reason is that, unlike the case of 3-dim, real Lie Algebras,
a unique decomposition of the structure constants' tensor in terms
of lower rank objects, has not been found. As a result, the
presentations of Petrov \cite{Petrov}, MacCallum \cite{MacCallum}
and Patera et. al. \cite{Patera}, differ substantially, especially
as far as the number of different real Lie Algebras, is concerned.

In the following the non vanishing structure constants for the
various 4-dimensional, real, Lie Algebras (according to Patera et.
al. Ref. \cite{Patera} which is considered to be the most complete
and extensive), the automorphism matrices and their generators,
are given in Table 1. Also, for each algebra, an irreducible form
of a generic, $4\times 4$, symmetric, positive definite, real
matrix is given, in Table 2, along with a suggested basis of
invariants.

We now come to the invariant description of a homogeneous 4-space.
Let $\sigma^{\alpha}_{i}(x)$ denote the basis of one forms,
invariant under the action of the symmetry group of motions,
acting simply transitively on the space. Then:
\begin{eqnarray}
\sigma^{\alpha}_{i,j}(x)-\sigma^{\alpha}_{j,i}(x)=
2C^{\alpha}_{\mu\nu}\sigma^{\mu}_{j}(x)\sigma^{\nu}_{i}(x)
\end{eqnarray}
where $C^{\alpha}_{\mu\nu}$, are the structure constants of the
corresponding Lie Algebra. Using this basis we can write, in these
coordinates the most general, manifestly invariant, line element
as:
\begin{eqnarray}
ds^{2}=\gamma_{\alpha\beta}\sigma^{\alpha}_{i}(x)\sigma^{\beta}_{j}(x)dx^{i}dx^{j}
\end{eqnarray}
where $\gamma_{\alpha\beta}$ is a numerical, $4\times 4$, real,
positive definite, symmetric matrix. If we consider the class of
general co-ordinate transformations (GCT's) $x^{i}=f^{i}(y^{m})$ which leave the given basis one forms
quasi-form invariant, i.e. those satisfying:
\begin{eqnarray} \label{Lambdadefinition}
\sigma^{\alpha}_{i}(x)\frac{\partial x^{i}}{\partial
y^{m}}=\Lambda^{\alpha}_{\mu}\sigma^{\mu}_{m}(y)
\end{eqnarray}
then we have a well defined, non trivial action, on the
configuration space, spanned by $\gamma_{\alpha\beta}$'s, given
by:
\begin{eqnarray} \label{gammalambda}
\widetilde{\gamma}_{\mu\nu}=\Lambda^{\alpha}_{\mu}\Lambda^{\beta}_{\nu}\gamma_{\alpha\beta}
\end{eqnarray}
The relevant result for 3-spaces, are given in \cite{ChrisCMP} and
the generalization to 4-spaces, is obvious. The requirement for
$\Lambda^{\alpha}_{\beta}$ to be constant leads, through the
integrability conditions for (\ref{Lambdadefinition}), to the
restrictions:
\begin{eqnarray} \label{AUT}
C^{\rho}_{\mu\nu}\Lambda^{\alpha}_{\rho}=\Lambda^{\kappa}_{\mu}\Lambda^{\sigma}_{\nu}C^{\alpha}_{\kappa\sigma}
\end{eqnarray}
which reveal $\Lambda^{\alpha}_{\beta}$, as an element of the
automorphism group of the corresponding Lie Algebra. Thus, the
configuration space, is divided into equivalence classes by the
action of the automorphism group according to (\ref{gammalambda}).

In order to have an infinitesimal description of this action, we
need to consider the generators $\lambda^{\alpha}_{\beta}$ of
$\Lambda^{\alpha}_{\beta}$. Their defining relations are
(\ref{generators}). We can easily see that the linear vector
fields in the configuration space:
\begin{eqnarray}
X_{(i)}=\lambda^{\alpha}_{\mu(i)}\gamma_{\alpha\nu}\frac{\partial}{\partial\gamma_{\mu\nu}}
\end{eqnarray}
induce, through their integral curves, exactly the motions
(\ref{gammalambda}) -- $(i)$ is a collective index corresponding
to a choice of base for $\lambda^{\alpha}_{\mu}$ and counts the
number of independent vector fields. Thus, if we wish for a scalar
function $\Psi=\Psi(\gamma_{\alpha\beta})$ to change only when we
move from one class to another, then we must demand:
\begin{eqnarray} \label{XPsi}
X_{(i)}\Psi=0, ~~~ \forall~~~i \in [1,\ldots, d], ~~~d<10
\end{eqnarray}
The solutions to this system of equations, say
$q^{A}=q^{A}(C^{\alpha}_{\mu\nu},\gamma_{\mu\nu})$ lead to the
finite description of the action of automorphisms -- with $A$
taking its values on the interval $[1,\ldots,10-d]$. By
construction, they satisfy:
\begin{eqnarray}
q^{(1)A}=q^{(2)A}~~~\textrm{for
every}~~~(\gamma^{(1)}_{\alpha\beta},\gamma^{(2)}_{\alpha\beta})
\end{eqnarray}
connected through an automorphism, as in (\ref{gammalambda}). A
kind of inverse to this proposition, which completes the finite
description of the 4-spaces, discussed here, is contained in the
following:
\vspace{0.5cm}
\noindent
\begin{thm}
Let $\gamma^{(1)}_{\alpha\beta},\gamma^{(2)}_{\alpha\beta}$ belong
to the configuration space. If $q^{(1)A}=q^{(2)A}$ $\forall ~A$,
then there is $\Lambda^{\alpha}_{\beta} \in Aut(G)$ such that
$\gamma^{(2)}_{\mu\nu}=\Lambda^{\alpha}_{\mu}
\Lambda^{\beta}_{\nu}\gamma^{(1)}_{\alpha\beta}$.
\end{thm}
\begin{proof}
We first observe that, as seen in Table 2, for each and every Lie
Algebra, the -- connected to the identity -- component of the
automorphism group, suffices to bring the generic, positive
definite, real, $4\times 4$, $\gamma_{\alpha\beta}$ to an
irreducible (though not unique) form
$\gamma^{~\textrm{Ir.}}_{\alpha\beta}$, possessing a number of
remaining arbitrary components which equals the number of
independent $q^{(A)}$'s. Thus, in this `'gauge`' the hypothesis of
the theorem i.e. $q^{(1)A}=q^{(2)A}$ $\forall ~A$, implies --
through the implicit function theorem \cite{VC} -- that
$\gamma^{~\textrm{Ir.}~(2)}_{\alpha\beta}=\gamma^{~\textrm{Ir.}~(1)}_{\alpha\beta}$.
Therefore, if
$\Lambda^{(2)\alpha}_{\beta},\Lambda^{(1)\alpha}_{\beta}$ are the
simplifying automorphisms, i.e.
$\gamma^{(2)}_{\mu\nu}=\Lambda^{(2)\alpha}_{\mu}
\Lambda^{(2)\beta}_{\nu}\gamma^{~\textrm{Ir.}~(2)}_{\alpha\beta}$
and $\gamma^{(1)}_{\mu\nu}=\Lambda^{(1)\alpha}_{\mu}
\Lambda^{(1)\beta}_{\nu}\gamma^{~\textrm{Ir.}~(1)}_{\alpha\beta}$
then the transformation\\
$\Lambda^{\alpha}_{\beta}=(\Lambda^{-1})^{(1)\alpha}_{\rho}\Lambda^{(2)\rho}_{\beta}$
connects $\gamma^{(2)}_{\mu\nu}$ to $\gamma^{(1)}_{\mu\nu}$ and
obviously belongs to $Aut(G)$.
\end{proof}

\vspace{0.5cm} Returning to the form of the solutions to equations
(\ref{XPsi}) it is straightforward to check that every scalar
combination of $C^{\lambda}_{\mu\nu}$ and $\gamma_{\mu\nu},
\gamma^{\alpha\beta}$, is annihilated by all $X_{(i)}$'s. The
number of independent such scalar contractions is, at most, six:
the 10 $\gamma_{\mu\nu}$'s plus the 12 $C^{\lambda}_{\mu\nu}$'s
(24 initially independent - 12 independent Jacobi Identities)
minus 16 arbitrary elements of $GL(4,\mathbb{R})$. The same number
is obtained by observing that the automorphism group always
contains the inner automorphism subgroup which has 4 generators
thus, there will be at most, 10-4=6 independent scalar
combinations.

A common, though not unique, suitable basis in the space of all
such scalars, valid for all 4-dim, Real Lie Algebras is:
\begin{subequations}
\begin{eqnarray}
q^{1},q^{2},q^{3},q^{4},q^{5},q^{6}
\end{eqnarray}
where:
\begin{eqnarray} \label{basis}
q^{1}&=&\Pi_{\alpha\beta\mu\nu}\gamma^{\alpha\mu}\gamma^{\beta\nu}\\
q^{2}&=&C^{\alpha}_{\beta\kappa}C^{\beta}_{\alpha\lambda}\gamma^{\kappa\lambda}\\
q^{3}&=&\Pi_{\alpha\beta\mu\nu}\Pi^{\alpha\beta\mu\nu}\\
q^{4}&=&\Pi_{\alpha\beta\kappa\lambda}\Pi_{\mu\nu\rho\sigma}\Pi^{\alpha\beta\mu\rho}\gamma^{\kappa\nu}\gamma^{\lambda\sigma}\\
q^{5}&=&\Upsilon_{\alpha}\Upsilon_{\beta}\gamma^{\alpha\beta}\\
q^{6}&=&\Omega_{\alpha}\Omega_{\beta}\gamma^{\alpha\beta}
\end{eqnarray}
\end{subequations}
with the allocations:
\begin{eqnarray}
\Pi_{\alpha\beta\mu\nu}&=&C^{\rho}_{\alpha\beta}C^{\sigma}_{\mu\nu}\gamma_{\rho\sigma}\\
\Upsilon_{\alpha}&=&\Pi_{\alpha\beta\mu\nu}C^{\nu}_{\kappa\lambda}\gamma^{\beta\lambda}\gamma^{\mu\kappa}\\
\Omega_{\alpha}&=&\Pi_{\alpha\beta\mu\nu}C^{\nu}_{\kappa\lambda}\Pi^{\beta\lambda\mu\kappa}
\end{eqnarray}
(Greek indices are raised and lowered with $\gamma^{\alpha\beta}$
and $\gamma_{\alpha\beta}$ respectively).

The number of functionally independent $q^{A}$'s is 6, only for a
5 out of the 30 homogeneous 4-spaces, here considered -- a fact
that is reminiscent of the analogous situation in Bianchi Types,
where only Type $VIII$ and $IX$ possess three functionally
independent $q$'s. For the rest of the cases, the number of
functionally independent $q^{A}$'s is less than 6.

These $q^{A}$'s can serve as `'coordinates`' of the reduced
configuration space on which the Wheeler-DeWitt equation, is to be
founded.

\section{Discussion}
We have investigated the action of the automorphism group of all
4-dim, Real Lie Algebras, in the space of $4\times 4$, real,
symmetric, and positive definite matrices, which is the
configuration space of homogeneous 4-spaces (admitting an
invariant basis of one-forms). These automorphisms naturally
emerge as the non trivial action of the Diffeomorphism Group on
this space. The finite invariant description of these 4-spaces, is
given in terms of the scalar combinations of the structure
constants with the scale factor matrix. The number of independent
such scalars is found to be at most 6. These scalars can be
considered either as the independent solutions to (\ref{XPsi}) or
as the differential scalar contractions constructed out of the
metric and its derivatives.\\
In 3-dim these are the curvature and/or higher derivative
curvature scalars; in 4-dim there are also scalar combinations of
the metric and its derivatives of degree greater than 3 (metric
invariants \cite{invar}), which can not be
expressed as higher derivative curvature invariants.\\
The $q^{A}$'s given in (\ref{basis}) may be curvature, higher
derivative curvature, or metric invariants. They irreducibly
characterize the corresponding homogeneous space. These quantities
are useful both at the classical and quantum level: classically
the irreducible of the scale factor matrix, (which in the light of
the Theorem are essentially the $q^{A}$'s) can be used as a
starting point for solving the Einstein Field Equations of the
corresponding 5-dim cosmological models. Quantum mechanically the
Wheeler-DeWitt equation, when an action principle exists, is to be
constructed on the reduced configuration space spanned by the
$q^{A}$'s. The reasoning of this is the desire to have ``gauge''
invariant wave functions.

Our analysis does not treat the Kantowski-Sacks like geometries
(homogeneous spaces with multiply transitive groups of motions).
These are 6 in number \cite{ishihara}.

Finally, we would like to mention a word about the computation of
the basis of invariants: it can be carried out using a symbolic
algebra package (such as Mathematica) and the hard thing is to
find a basis valid for all the 30 homogeneous spaces.The price one
pays is that one has to consider as many as 10 powers of the
structure constants ( i.e. 10 derivatives of the metric . This,
however,is a worthwhile sacrifice of simplicity, since it will
enable one to make comparative studies of the corresponding
5-dimensional Quantum Cosmologies.
\newpage
\section*{Table Captions}
\textbf{Table1}\\
The first column gives the names of the Lie Algebras according to \cite{Patera}.
The second, gives the non vanishing structure constants. Column three, exhibits
the corresponding automorphism group $\Lambda^{\alpha}_{\beta}$ --along with the
disconnected to the identity component, when it exists. Finally, in column four,
the generators of the automorphism group $\lambda^{\alpha}_{\beta}$, are presented.

\noindent
\textbf{Table2}\\
The first column gives the names of the Lie Algebras according to \cite{Patera}.
The second, gives a possible irreducible form for a generic $4 \times 4$, real,
symmetric and positive definite matrix, $\gamma_{\alpha\beta}$ as resulting by the
action of the corresponding automorphism group. Care has been taken, so that
the exhibited reduced form can always be achieved. Finally, in column three, a
set of functionally idependent metric invariants, is given. Functional independence
has been tested using the reduced form of $\gamma_{\alpha\beta}$'s --given
in column 2. However, since the reduction has been achieved by the appropriate
automorphism group, and since this action is nothing but the effect of a
general co-ordinate transformation, one concludes that the given metric invariants
are functionally indepedendet, with respect to the generic $\gamma_{\alpha\beta}$,
as well.

\newpage
\landscape
\begin{center}
\begin{tabular}{|c|c|c|c|}
\multicolumn{4}{c}{TABLE 1}\\
  \hline
  \hline
  Lie Algebra & Non Vanishing & Automorphisms & Generators \\
              &Structure Constants&$\Lambda^{\alpha}_{\beta}$&$\lambda^{\alpha}_{\beta}$\\
  \hline
  \hline
  $4A_{1}$ &  & $GL(4,\Re)$ & $GL(4,\Re)$ \\
  \hline
  $A_{2}\oplus A_{1}$ & $C^{2}_{12}=1$ & $\left(
\begin{array}{cccc}
  1 & 0 & 0 & 0 \\
  a_{5} & a_{6} & 0 & 0 \\
  a_{9} & 0 & a_{11} & a_{12} \\
  a_{13} & 0 & a_{15} & a_{16} \\
\end{array}
\right)$ & $\left(
\begin{array}{cccc}
  0 & 0 & 0 & 0 \\
  g_{5} & g_{6} & 0 & 0 \\
  g_{9} & 0 & g_{11} & g_{12} \\
  g_{13} & 0 & g_{15} & g_{16} \\
\end{array}
\right)$ \\
  \hline
  $2A_{2}$ & $C^{2}_{12}=1$ $C^{4}_{34}=1$ & $\left(
\begin{array}{cccc}
  1 & 0 & 0 & 0 \\
  a_{5} & a_{6} & 0 & 0 \\
  0 & 0 & 1 & 0 \\
  0 & 0 & a_{15} & a_{16} \\
\end{array}
\right)$ \textrm{or} $\left(
\begin{array}{cccc}
  0 & 0 & 1 & 0 \\
  0 & 0 & a_{7} & a_{8} \\
  1 & 0 & 0 & 0 \\
  a_{13} & a_{14} & 0 & 0 \\
\end{array}
\right)$ & $\left(
\begin{array}{cccc}
  0 & 0 & 0 & 0 \\
  g_{5} & g_{6} & 0 & 0 \\
  0 & 0 & 0 & 0 \\
  0 & 0 & g_{15} & g_{16} \\
\end{array}
\right)$ \\
\hline
  $A_{3,1}\oplus A_{1}$ & $C^{1}_{23}=1$ & $\left(
\begin{array}{cccc}
  a_{11}a_{6}-a_{10}a_{7} & a_{2} & a_{3} & a_{4} \\
  0 & a_{6} & a_{7} & 0 \\
  0 & a_{10} & a_{11} & 0 \\
  0 & a_{14} & a_{15} & a_{16} \\
\end{array}
\right)$ & $\left(
\begin{array}{cccc}
  g_{6}+g_{11} & g_{2} & g_{3} & g_{4} \\
  0 & g_{6} & g_{7} & 0 \\
  0 & g_{10} & g_{11} & 0 \\
  0 & g_{14} & g_{15} & g_{16} \\
\end{array}
\right)$ \\
  \hline
$A_{3,2}\oplus A_{1}$ & $C^{1}_{13}=1$ $C^{1}_{23}=1$
$C^{2}_{23}=1$ & $\left(
\begin{array}{cccc}
  a_{1} & a_{2} & a_{3} & 0 \\
  0 & a_{1} & a_{7} & 0 \\
  0 & 0 & 1 & 0 \\
  0 & 0 & a_{15} & a_{16} \\
\end{array}
\right)$ & $\left(
\begin{array}{cccc}
  g_{1} & g_{2} & g_{3} & 0 \\
  0 & g_{1} & g_{7} & 0 \\
  0 & 0 & 0 & 0 \\
  0 & 0 & g_{15} & g_{16} \\
\end{array}
\right)$ \\
\hline $A_{3,3}\oplus A_{1}$ & $C^{1}_{13}=1$ $C^{2}_{23}=1$ &
$\left(
\begin{array}{cccc}
  a_{1} & a_{2} & a_{3} & 0 \\
  a_{5} & a_{6} & a_{7} & 0 \\
  0 & 0 & 1 & 0 \\
  0 & 0 & a_{15} & a_{16} \\
\end{array}
\right)$ & $\left(
\begin{array}{cccc}
  g_{1} & g_{2} & g_{3} & 0 \\
  g_{5} & g_{6} & g_{7} & 0 \\
  0 & 0 & 0 & 0 \\
  0 & 0 & g_{15} & g_{16} \\
\end{array}
\right)$ \\
\hline $A_{3,4}\oplus A_{1}$ & $C^{1}_{13}=1$ $C^{2}_{23}=-1$ &
$\left(
\begin{array}{cccc}
  a_{1} & 0 & a_{3} & 0 \\
  0 & a_{6} & a_{7} & 0 \\
  0 & 0 & 1 & 0 \\
  0 & 0 & a_{15} & a_{16} \\
\end{array}
\right)$ \textrm{or} $\left(
\begin{array}{cccc}
  0 & a_{2} & a_{3} & 0 \\
  a_{5} & 0 & a_{7} & 0 \\
  0 & 0 & -1 & 0 \\
  0 & 0 & a_{15} & a_{16} \\
\end{array}
\right)$ & $\left(
\begin{array}{cccc}
  g_{1} & 0 & g_{3} & 0 \\
  0 & g_{6} & g_{7} & 0 \\
  0 & 0 & 0 & 0 \\
  0 & 0 & g_{15} & g_{16} \\
\end{array}
\right)$ \\
\hline
\end{tabular}
\end{center}
\endlandscape
\landscape
\begin{center}
\begin{tabular}{|c|c|c|c|}
\multicolumn{4}{c}{TABLE 1 (Continued)}\\
  \hline
  \hline
  Lie Algebra & Non Vanishing & Automorphisms & Generators \\
              &Structure Constants&$\Lambda^{\alpha}_{\beta}$&$\lambda^{\alpha}_{\beta}$\\
  \hline
  \hline
  $A^{\alpha}_{3,5}\oplus A_{1}$ &  &  &  \\
  $0<|\alpha|<1$ & $C^{1}_{13}=1$ $C^{2}_{23}=\alpha$ & $\left(
\begin{array}{cccc}
  a_{1} & 0 & a_{3} & 0 \\
  0 & a_{6} & a_{7} & 0 \\
  0 & 0 & 1 & 0 \\
  0 & 0 & a_{15} & a_{16} \\
\end{array}
\right)$ & $\left(
\begin{array}{cccc}
  g_{1} & 0 & g_{3} & 0 \\
  0 & g_{6} & g_{7} & 0 \\
  0 & 0 & 0 & 0 \\
  0 & 0 & g_{15} & g_{16} \\
\end{array}
\right)$\\
  \hline
  $A_{3,6}\oplus A_{1}$ & $C^{2}_{13}=-1$ $C^{1}_{23}=1$ & $\left(
\begin{array}{cccc}
  a_{1} & a_{2} & a_{3} & 0 \\
  -a_{2} & a_{1} & a_{7} & 0 \\
  0 & 0 & 1 & 0 \\
  0 & 0 & a_{15} & a_{16} \\
\end{array}
\right)$ \textrm{,} $\left(
\begin{array}{cccc}
  a_{1} & a_{2} & a_{3} & 0 \\
  a_{2} & -a_{1} & a_{7} & 0 \\
  0 & 0 & -1 & 0 \\
  0 & 0 & a_{15} & a_{16} \\
\end{array}
\right)$ & $\left(
\begin{array}{cccc}
  g_{1} & g_{2} & g_{3} & 0 \\
  -g_{2} & g_{1} & g_{4} & 0 \\
  0 & 0 & 0 & 0 \\
  0 & 0 & g_{15} & g_{16} \\
\end{array}
\right)$ \\
  \hline
  $A^{\alpha}_{3,7}\oplus A_{1}$ & $C^{1}_{13}=\alpha$ $C^{2}_{13}=-1$ $C^{1}_{23}=1$ &  &  \\
  $0<\alpha$ & $C^{2}_{23}=\alpha$ & $\left(
\begin{array}{cccc}
  a_{1} & a_{2} & a_{3} & 0 \\
  -a_{2} & a_{1} & a_{7} & 0 \\
  0 & 0 & 1 & 0 \\
  0 & 0 & a_{15} & a_{16} \\
\end{array}
\right)$ & $\left(
\begin{array}{cccc}
  g_{1} & g_{2} & g_{3} & 0 \\
  -g_{2} & g_{1} & g_{7} & 0 \\
  0 & 0 & 0 & 0 \\
  0 & 0 & g_{15} & g_{16} \\
\end{array}
\right)$\\
  \hline
  $A_{3,8}\oplus A_{1}$ & $C^{1}_{23}=1$ $C^{2}_{13}=-1$ $C^{3}_{12}=-1$ & \multicolumn{1}{c}{see Appendix} &\\
  \hline
  $A_{3,9}\oplus A_{1}$ & $C^{3}_{12}=1$ $C^{1}_{23}=1$ $C^{2}_{31}=1$ & \multicolumn{1}{c}{see Appendix} &\\
  \hline
  $A_{4,1}$ & $C^{1}_{24}=1$ $C^{2}_{34}=1$ & $\left(
\begin{array}{cccc}
  a_{11}a^{2}_{16} & a_{7}a_{16} & a_{3} & a_{4} \\
  0 & a_{11}a_{16} & a_{7} & a_{8} \\
  0 & 0 & a_{11} & a_{12} \\
  0 & 0 & 0 & a_{16} \\
\end{array}
\right)$ & $\left(
\begin{array}{cccc}
  g_{11}+2g_{16} & g_{7} & g_{3} & g_{4} \\
  0 & g_{11}+g_{16} & g_{7} & g_{8} \\
  0 & 0 & g_{11} & g_{12} \\
  0 & 0 & 0 & g_{16} \\
\end{array}
\right)$ \\
  \hline
  \end{tabular}
\end{center}
\endlandscape
\landscape
\begin{center}
\begin{tabular}{|c|c|c|c|}
\multicolumn{4}{c}{TABLE 1 (Continued)}\\
  \hline
  \hline
  Lie Algebra & Non Vanishing & Automorphisms & Generators \\
              &Structure Constants&$\Lambda^{\alpha}_{\beta}$&$\lambda^{\alpha}_{\beta}$\\
  \hline
  \hline
  $A^{\alpha}_{4,2}$ & $C^{1}_{14}=\alpha$ $C^{2}_{24}=1$ $C^{2}_{34}=1$ &  &  \\
  $\alpha\neq(0,1)$ & $C^{3}_{34}=1$ & $\left(
\begin{array}{cccc}
  a_{1} & 0 & 0 & a_{4} \\
  0 & a_{6} & 0 & a_{8} \\
  0 & 0 & a_{6} & 0 \\
  0 & 0 & 0 & 1 \\
\end{array}
\right)$ & $\left(
\begin{array}{cccc}
  g_{1} & 0 & 0 & g_{4} \\
  0 & g_{6} & 0 & g_{8} \\
  0 & 0 & g_{6} & 0 \\
  0 & 0 & 0 & 0 \\
\end{array}
\right)$\\
  \hline
  $A^{1}_{4,2}$ & $C^{1}_{14}=1$ $C^{2}_{24}=1$ $C^{2}_{34}=1$ &  &  \\
  & $C^{3}_{34}=1$ & $\left(
\begin{array}{cccc}
  a_{1} & 0 & 0 & a_{4} \\
  a_{5} & a_{6} & 0 & a_{8} \\
  0 & 0 & a_{6} & 0 \\
  0 & 0 & 0 & 1 \\
\end{array}
\right)$ & $\left(
\begin{array}{cccc}
  g_{1} & 0 & 0 & g_{4} \\
  g_{5} & g_{11} & 0 & g_{8} \\
  0 & 0 & g_{11} & 0 \\
  0 & 0 & 0 & 0 \\
\end{array}
\right)$\\
\hline $A_{4,3}$ & $C^{1}_{14}=1$ $C^{2}_{34}=1$ & $\left(
\begin{array}{cccc}
  a_{1} & 0 & 0 & a_{4} \\
  0 & a_{6} & a_{7} & a_{8} \\
  0 & 0 & a_{6} & a_{12} \\
  0 & 0 & 0 & 1 \\
\end{array}
\right)$ & $\left(
\begin{array}{cccc}
  g_{1} & 0 & 0 & g_{4} \\
  0 & g_{11} & g_{7} & g_{8} \\
  0 & 0 & g_{11} & g_{12} \\
  0 & 0 & 0 & 0 \\
\end{array}
\right)$ \\
\hline
$A_{4,4}$ & $C^{1}_{14}=1$ $C^{1}_{24}=1$ $C^{2}_{24}=1$ &
&\\
 & $C^{2}_{34}=1$ $C^{3}_{34}=1$ & $\left(
\begin{array}{cccc}
  a_{1} & a_{2} & a_{3} & a_{4} \\
  0 & a_{1} & a_{2} & a_{8} \\
  0 & 0 & a_{1} & a_{12} \\
  0 & 0 & 0 & 1 \\
\end{array}
\right)$ & $\left(
\begin{array}{cccc}
  g_{1} & g_{7} & g_{3} & g_{4} \\
  0 & g_{1} & g_{7} & g_{8} \\
  0 & 0 & g_{1} & g_{12} \\
  0 & 0 & 0 & 0 \\
\end{array}
\right)$ \\
\hline
$A^{\alpha,\beta}_{4,5}$ & $C^{1}_{14}=1$
$C^{2}_{24}=\alpha$ $C^{3}_{34}=\beta$ &
&\\
$-1\leq\alpha<\beta<1$, $\alpha\beta\neq 0$ &  & $\left(
\begin{array}{cccc}
  a_{1} & 0 & 0 & a_{4} \\
  0 & a_{6} & 0 & a_{8} \\
  0 & 0 & a_{11} & a_{12} \\
  0 & 0 & 0 & 1 \\
\end{array}
\right)$ & $\left(
\begin{array}{cccc}
  g_{1} & 0 & 0 & g_{4} \\
  0 & g_{6} & 0 & g_{8} \\
  0 & 0 & g_{11} & g_{12} \\
  0 & 0 & 0 & 0 \\
\end{array}
\right)$ \\
\hline
\end{tabular}
\end{center}
\endlandscape
\landscape
\begin{center}
\begin{tabular}{|c|c|c|c|}
\multicolumn{4}{c}{TABLE 1 (Continued)}\\
  \hline
  \hline
  Lie Algebra & Non Vanishing & Automorphisms & Generators \\
              &Structure Constants&$\Lambda^{\alpha}_{\beta}$&$\lambda^{\alpha}_{\beta}$\\
  \hline
  \hline
$A^{\alpha,\alpha}_{4,5}$ & $C^{1}_{14}=1$ $C^{2}_{24}=\alpha$
$C^{3}_{34}=\alpha$ &
&\\
$-1\leq\alpha<1$, $\alpha\neq 0$ &  & $\left(
\begin{array}{cccc}
  a_{1} & 0 & 0 & a_{4} \\
  0 & a_{6} & a_{7} & a_{8} \\
  0 & a_{10} & a_{11} & a_{12} \\
  0 & 0 & 0 & 1 \\
\end{array}
\right)$ & $\left(
\begin{array}{cccc}
  g_{1} & 0 & 0 & g_{4} \\
  0 & g_{6} & g_{7} & g_{8} \\
  0 & g_{10} & g_{11} & g_{12} \\
  0 & 0 & 0 & 0 \\
\end{array}
\right)$ \\
\hline
$A^{\alpha,1}_{4,5}$ & $C^{1}_{14}=1$ $C^{2}_{24}=\alpha$
$C^{3}_{34}=1$ &
&\\
$-1\leq\alpha<1$, $\alpha\neq 0$ &  & $\left(
\begin{array}{cccc}
  a_{1} & 0 & a_{3} & a_{4} \\
  0 & a_{6} & 0 & a_{8} \\
  a_{9} & 0 & a_{11} & a_{12} \\
  0 & 0 & 0 & 1 \\
\end{array}
\right)$ & $\left(
\begin{array}{cccc}
  g_{1} & 0 & g_{3} & g_{4} \\
  0 & g_{6} & 0 & g_{8} \\
  g_{9} & 0 & g_{11} & g_{12} \\
  0 & 0 & 0 & 0 \\
\end{array}
\right)$ \\
\hline $A^{1,1}_{4,5}$ & $C^{1}_{14}=1$ $C^{2}_{24}=1$
$C^{3}_{34}=1$ & $\left(
\begin{array}{cccc}
  a_{1} & a_{2} & a_{3} & a_{4} \\
  a_{5} & a_{6} & a_{7} & a_{8} \\
  a_{9} & a_{10} & a_{11} & a_{12} \\
  0 & 0 & 0 & 1 \\
\end{array}
\right)$ & $\left(
\begin{array}{cccc}
  g_{1} & g_{2} & g_{3} & g_{4} \\
  g_{5} & g_{6} & g_{7} & g_{8} \\
  g_{9} & g_{10} & g_{11} & g_{12} \\
  0 & 0 & 0 & 0 \\
\end{array}
\right)$\\
\hline
$A^{\alpha,\beta}_{4,6}$ & $C^{1}_{14}=\alpha$
$C^{2}_{24}=\beta$ $C^{3}_{24}=-1$ &
&\\
$\alpha\neq 0$, $\beta\geq 0$ & $C^{2}_{34}=1$ $C^{3}_{34}=\beta$
& $\left(
\begin{array}{cccc}
  a_{1} & 0 & 0 & a_{4} \\
  0 & a_{6} & a_{7} & a_{8} \\
  0 & -a_{7} & a_{6} & a_{12} \\
  0 & 0 & 0 & 1 \\
\end{array}
\right)$ & $\left(
\begin{array}{cccc}
  g_{1} & 0 & 0 & g_{4} \\
  0 & g_{11} & -g_{10} & g_{8} \\
  0 & g_{10} & g_{11} & g_{12} \\
  0 & 0 & 0 & 0 \\
\end{array}
\right)$ \\
\hline
$A_{4,7}$ & $C^{1}_{14}=2$ $C^{2}_{24}=1$ $C^{2}_{34}=1$ & &\\
& $C^{3}_{34}=1$ $C^{1}_{23}=1$ & $\left(
\begin{array}{cccc}
  a^{2}_{6} & -a_{12}a_{6} & -a_{12}(a_{6}+a_{7})+a_{6}a_{8} & a_{4} \\
  0 & a_{6} & a_{7} & a_{8} \\
  0 & 0 & a_{6} & a_{12} \\
  0 & 0 & 0 & 1 \\
\end{array}
\right)$ & $\left(
\begin{array}{cccc}
  2g_{11} & -g_{12} & -g_{12}+g_{8} & g_{4} \\
  0 & g_{11} & g_{7} & g_{8} \\
  0 & 0 & g_{11} & g_{12} \\
  0 & 0 & 0 & 0 \\
\end{array}
\right)$\\
\hline
\end{tabular}
\end{center}
\endlandscape
\landscape
\begin{center}
\begin{tabular}{|c|c|c|c|}
\multicolumn{4}{c}{TABLE 1 (Continued)}\\
  \hline
  \hline
  Lie Algebra & Non Vanishing & Automorphisms & Generators \\
              &Structure Constants&$\Lambda^{\alpha}_{\beta}$&$\lambda^{\alpha}_{\beta}$\\
  \hline
  \hline
$A_{4,8}$ & $C^{1}_{23}=1$ $C^{2}_{24}=1$ $C^{3}_{34}=-1$ &
$\left(
\begin{array}{cccc}
  a_{11}a_{6} & a_{12}a_{6} & a_{11}a_{8} & a_{4} \\
  0 & a_{6} & 0 & a_{8} \\
  0 & 0 & a_{11} & a_{12} \\
  0 & 0 & 0 & 1 \\
\end{array}
\right)$ \textrm{or} & $\left(
\begin{array}{cccc}
  g_{11}+g_{6} & g_{12} & g_{8} & g_{4} \\
  0 & g_{6} & 0 & g_{8} \\
  0 & 0 & g_{11} & g_{12} \\
  0 & 0 & 0 & 0 \\
\end{array}
\right)$ \\
& & $\left(
\begin{array}{cccc}
  -a_{10}a_{7} & -a_{10}a_{8} & -a_{12}a_{7} & a_{4} \\
  0 & 0 & a_{7} & a_{8} \\
  0 & a_{10} & 0 & a_{12} \\
  0 & 0 & 0 & -1 \\
\end{array}
\right)$ & \\
\hline
$A^{\beta}_{4,9}$ & $C^{1}_{23}=1$ $C^{1}_{14}=1+\beta$
$C^{2}_{24}=1$ &
&\\
$0<|\beta|<1$ &  $C^{3}_{34}=\beta$ & $\left(
\begin{array}{cccc}
  a_{11}a_{6} & -a_{12}a_{6}/\beta & a_{8}a_{11} & a_{4} \\
  0 & a_{6} & 0 & a_{8} \\
  0 & 0 & a_{11} & a_{12} \\
  0 & 0 & 0 & 1 \\
\end{array}
\right)$ & $\left(
\begin{array}{cccc}
  g_{11}+g_{6} & -g_{12}/\beta & g_{8} & g_{4} \\
  0 & g_{6} & 0 & g_{8} \\
  0 & 0 & g_{11} & g_{12} \\
  0 & 0 & 0 & 0 \\
\end{array}
\right)$ \\
\hline
$A^{1}_{4,9}$ & $C^{1}_{23}=1$ $C^{1}_{14}=2$
$C^{2}_{24}=1$ &
&\\
 &  $C^{3}_{34}=1$ & $\left(
\begin{array}{cccc}
  a_{11}a_{6}-a_{10}a_{7} & -a_{12}a_{6}+a_{10}a_{8} & a_{8}a_{11}-a_{7}a_{12} & a_{4} \\
  0 & a_{6} & a_{7} & a_{8} \\
  0 & a_{10} & a_{11} & a_{12} \\
  0 & 0 & 0 & 1 \\
\end{array}
\right)$ & $\left(
\begin{array}{cccc}
  g_{11}+g_{6} & -g_{12} & g_{8} & g_{4} \\
  0 & g_{6} & g_{7} & g_{8} \\
  0 & g_{10} & g_{11} & g_{12} \\
  0 & 0 & 0 & 0 \\
\end{array}
\right)$ \\
\hline $A^{0}_{4,9}$ & $C^{1}_{23}=1$ $C^{1}_{14}=1$
$C^{2}_{24}=1$ & $\left(
\begin{array}{cccc}
  a_{11}a_{6} & a_{2} & a_{8}a_{11} & a_{4} \\
  0 & a_{6} & 0 & a_{8} \\
  0 & 0 & a_{11} & 0 \\
  0 & 0 & 0 & 1 \\
\end{array}
\right)$ & $\left(
\begin{array}{cccc}
  g_{11}+g_{6} & g_{2} & g_{8} & g_{4} \\
  0 & g_{6} & 0 & g_{8} \\
  0 & 0 & g_{11} & 0 \\
  0 & 0 & 0 & 0 \\
\end{array}
\right)$\\
\hline
\end{tabular}
\end{center}
\endlandscape
\landscape
\begin{center}
\begin{tabular}{|c|c|c|c|}
\multicolumn{4}{c}{TABLE 1 (Continued)}\\
  \hline
  \hline
  Lie Algebra & Non Vanishing & Automorphisms & Generators \\
              &Structure Constants&$\Lambda^{\alpha}_{\beta}$&$\lambda^{\alpha}_{\beta}$\\
  \hline
  \hline
$A_{4,10}$ & $C^{1}_{23}=1$ $C^{3}_{24}=-1$ $C^{2}_{34}=1$ &
$\left(
\begin{array}{cccc}
  a^{2}_{6}+a^{2}_{7} & a_{12}a_{7}-a_{6}a_{8} & -a_{12}a_{6}-a_{7}a_{8} & a_{4} \\
  0 & a_{6} & a_{7} & a_{8} \\
  0 & -a_{7} & a_{6} & a_{12} \\
  0 & 0 & 0 & 1 \\
\end{array}
\right)$ \textrm{or} & $\left(
\begin{array}{cccc}
  2g_{11} & -g_{8} & -g_{12} & g_{4} \\
  0 & g_{11} & -g_{10} & g_{8} \\
  0 & g_{10} & g_{11} & g_{12} \\
  0 & 0 & 0 & 0 \\
\end{array}
\right)$
\\
& & $\left(
\begin{array}{cccc}
  -a^{2}_{6}-a^{2}_{7} & a_{12}a_{7}+a_{6}a_{8} & -a_{12}a_{6}+a_{7}a_{8} & a_{4} \\
  0 & a_{6} & a_{7} & a_{8} \\
  0 & a_{7} & -a_{6} & a_{12} \\
  0 & 0 & 0 & -1 \\
\end{array}
\right)$ & \\
\hline
$A^{\alpha}_{4,11}$ & $C^{1}_{23}=1$ $C^{1}_{14}=2\alpha$
$C^{2}_{24}=\alpha$ & &\\
$\alpha>0$ & $C^{3}_{24}=-1$ $C^{2}_{34}=1$ $C^{3}_{34}=\alpha$ &
$\left(
\begin{array}{cccc}
  a^{2}_{6}+a^{2}_{7} & -(w1)/(1+\alpha^{2}) & -(w2)/(1+\alpha^{2}) & a_{4} \\
  0 & a_{6} & a_{7} & a_{8} \\
  0 & -a_{7} & a_{6} & a_{12} \\
  0 & 0 & 0 & 1 \\
\end{array}
\right)$ & $\left(
\begin{array}{cccc}
  2g_{11} & g_{2} & g_{3} & g_{4} \\
  0 & g_{11} & -g_{10} & -g_{2}+\alpha g_{3} \\
  0 & g_{10} & g_{11} & -\alpha g_{2}-g_{3} \\
  0 & 0 & 0 & 0\\
\end{array}
\right)$ \\
 & & $w1=a_{6}(\alpha ~a_{12}+a_{8})+a_{7}(\alpha ~a_{8}-a_{12})$ & \\
 & & $w2=a_{6}(a_{12}-\alpha ~a_{8})+a_{7}(\alpha ~a_{12}+a_{8})$ & \\
\hline $A_{4,12}$ & $C^{1}_{13}=1$ $C^{2}_{23}=1$
$C^{2}_{14}=-1$ & &\\
& $C^{1}_{24}=1$ & $\left(
\begin{array}{cccc}
  a_{1} & a_{2} & a_{3} & a_{4} \\
  -a_{2} & a_{1} & a_{4} & -a_{3} \\
  0 & 0 & 1 & 0 \\
  0 & 0 & 0 & 1 \\
\end{array}
\right)$ \textrm{or} $\left(
\begin{array}{cccc}
  a_{1} & a_{2} & a_{3} & a_{4} \\
  a_{2} & -a_{1} & -a_{4} & a_{3} \\
  0 & 0 & 1 & 0 \\
  0 & 0 & 0 & -1 \\
\end{array}
\right)$ & $\left(
\begin{array}{cccc}
  g_{6} & -g_{5} & -g_{8} & g_{7} \\
  g_{5} & g_{6} & g_{7} & g_{8} \\
  0 & 0 & 0 & 0\\
  0 & 0 & 0 & 0\\
\end{array}
\right)$ \\
\hline
\end{tabular}
\end{center}
\endlandscape
\appendix
\section{Appendix}
We give the automorphisms and their generators for $A_{3,8}\oplus
A_{1}$ and $A_{3,9}\oplus A_{1}$ Lie Algebras, in symbolic
(matrix) notation.
\begin{itemize}
\item[$A_{3,8}\oplus A_{1}$]
\begin{eqnarray}
\Lambda=\textrm{Rotation}_{xy}\textrm{Boost}_{xz}\textrm{Boost}_{yz}C
\end{eqnarray}
where:
\begin{eqnarray}
\textrm{Rotation}_{xy}&=&\left(
\begin{array}{cccc}
   \cos(a_{1}) & \sin(a_{1}) & 0 & 0 \\
   -\sin(a_{1}) & \cos(a_{1}) & 0 & 0 \\
   0 & 0 & 1 & 0 \\
   0 & 0 & 0 & 1
\end{array}
\right)\\
\textrm{Boost}_{xz}&=&\left(
\begin{array}{cccc}
  \cosh(a_{2}) & 0 & \sinh(a_{2}) & 0\\
   0 & 1 & 0 & 0 \\
  \sinh(a_{2}) & 0 & \cosh(a_{2}) & 0 \\
   0 & 0 & 0 & 1
\end{array}
\right)\\
\textrm{Boost}_{yz}&=&\left(
\begin{array}{cccc}
   1 & 0 & 0 & 0\\
   0 & \cosh(a_{3}) & \sinh(a_{3}) & 0 \\
   0 & \sinh(a_{3}) & \cosh(a_{3}) & 0 \\
   0 & 0 & 0 & 1
\end{array}
\right)\\
C&=&\left(
\begin{array}{cccc}
  1 & 0 & 0 & 0\\
  0 & 1 & 0 & 0 \\
  0 & 0 & 1 & 0 \\
  0 & 0 & 0 & a_{4}
\end{array}
\right)
\end{eqnarray}
The generator:
\begin{eqnarray}
\left(
\begin{array}{cccc}
   0 & g_{2} & g_{3} & 0 \\
   -g_{2} & 0 & g_{7} & 0 \\
   g_{3} & g_{7} & 0 & 0 \\
   0 & 0 & 0 & g_{16}
\end{array}
\right)
\end{eqnarray}
\newpage
\item[$A_{3,9}\oplus A_{1}$]
\begin{eqnarray}
\Lambda=\textrm{Rotation}_{xy}\textrm{Rotation}_{xz}\textrm{Rotation}_{yz}C
\end{eqnarray}
where:
\begin{eqnarray}
\textrm{Rotation}_{xy}&=&\left(
\begin{array}{cccc}
   \cos(a_{1}) & \sin(a_{1}) & 0 & 0 \\
   -\sin(a_{1}) & \cos(a_{1}) & 0 & 0 \\
   0 & 0 & 1 & 0 \\
   0 & 0 & 0 & 1
\end{array}
\right)\\
\textrm{Rotation}_{xz}&=&\left(
\begin{array}{cccc}
  \cos(a_{2}) & 0 & -\sin(a_{2}) & 0\\
   0 & 1 & 0 & 0 \\
  \sin(a_{2}) & 0 & \cos(a_{2}) & 0 \\
   0 & 0 & 0 & 1
\end{array}
\right)\\
\textrm{Rotation}_{yz}&=&\left(
\begin{array}{cccc}
   1 & 0 & 0 & 0\\
   0 & \cos(a_{3}) & \sin(a_{3}) & 0 \\
   0 & -\sin(a_{3}) & \cos(a_{3}) & 0 \\
   0 & 0 & 0 & 1
\end{array}
\right)\\
C&=&\left(
\begin{array}{cccc}
  1 & 0 & 0 & 0\\
  0 & 1 & 0 & 0 \\
  0 & 0 & 1 & 0 \\
  0 & 0 & 0 & a_{4}
\end{array}
\right)
\end{eqnarray}
The generator:
\begin{eqnarray}
\left(
\begin{array}{cccc}
   0 & g_{2} & g_{3} & 0 \\
   -g_{2} & 0 & g_{7} & 0 \\
   -g_{3} & -g_{7} & 0 & 0 \\
   0 & 0 & 0 & g_{16}
\end{array}
\right)
\end{eqnarray}
\end{itemize}
\newpage
\landscape
\begin{center}
\begin{tabular}{|c|c|c|}
\multicolumn{3}{c}{TABLE 2}\\
  \hline
  \hline
  Lie Algebra & (Possible) Reducible Form of $\gamma_{\alpha\beta}$ & Functionally Independent Invariants \\
  \hline
  \hline
  $4A_{1}$ & $\left(\begin{array}{cccc}
    1 & 0 & 0 & 0 \\
    0 & 1 & 0 & 0 \\
    0 & 0 & 1 & 0 \\
    0 & 0 & 0 & 1 \\
  \end{array}\right)$ & None \\
  \hline
  $A_{2}\oplus 2A_{1}$ & $\left(\begin{array}{cccc}
    \gamma_{11} & 0 & 0 & 0 \\
    0 & 1 & 0 & \gamma_{24} \\
    0 & 0 & 1 & 0 \\
    0 & \gamma_{24} & 0 & 1\\
  \end{array}\right)$ & $q^{1},q^{2}$ \\
  \hline
  $2A_{2}$ & $\left(\begin{array}{cccc}
    \gamma_{11} & 0 & \gamma_{13} & \gamma_{14} \\
    0 & 1 & \gamma_{23} & \gamma_{24} \\
    \gamma_{13} & \gamma_{23} & \gamma_{33} & 0 \\
    \gamma_{14} & \gamma_{24} & 0 & 1 \\
  \end{array}\right)$ & $q^{1},q^{2},q^{3},q^{4},q^{5},q^{6}$ \\
  \hline
  $A_{3,1}\oplus A_{1}$ & $\left(\begin{array}{cccc}
    \gamma_{11} & 0 & 0 & 0 \\
    0 & 1 & 0 & 0 \\
    0 & 0 & 1 & 0 \\
    0 & 0 & 0 & 1 \\
  \end{array}\right)$ & $q^{1}$ \\
  \hline
  $A_{3,2}\oplus A_{1}$ & $\left(\begin{array}{cccc}
    1 & 0 & 0 & \gamma_{14} \\
    0 & \gamma_{22} & 0 & \gamma_{24}\\
    0 & 0 & \gamma_{33} & 0 \\
    \gamma_{14} & \gamma_{24} & 0 & 1 \\
  \end{array}\right)$ & $q^{1},q^{2},q^{3},q^{5}$ \\
  \hline
  $A_{3,3}\oplus A_{1}$ & $\left(\begin{array}{cccc}
    1 & 0 & 0 & \gamma_{14} \\
    0 & 1 & 0 & 0\\
    0 & 0 & \gamma_{33} & 0 \\
    \gamma_{14} & 0 & 0 & 1 \\
  \end{array}\right)$ & $q^{1},q^{2}$ \\
  \hline
  $A_{3,4}\oplus A_{1}$ & $\left(\begin{array}{cccc}
    1 & \gamma_{12} & 0 & \gamma_{14} \\
    \gamma_{12} & 1 & 0 & \gamma_{24}\\
    0 & 0 & \gamma_{33} & 0 \\
    \gamma_{14} & \gamma_{24} & 0 & 1 \\
  \end{array}\right)$ & $q^{1},q^{2},q^{3},q^{5}$ \\
  \hline
\hline
\end{tabular}
\end{center}
\endlandscape
\newpage
\landscape
\begin{center}
\begin{tabular}{|c|c|c|}
\multicolumn{3}{c}{TABLE 2 (Continued)}\\
  \hline
  \hline
  Lie Algebra & (Possible) Reducible Form of $\gamma_{\alpha\beta}$ & Functionally Independent Invariants \\
  \hline
  \hline
  $A^{\alpha}_{3,5}\oplus A_{1}$ & $\left(\begin{array}{cccc}
    1 & \gamma_{12} & 0 & \gamma_{14} \\
    \gamma_{12} & 1 & 0 & \gamma_{24}\\
    0 & 0 & \gamma_{33} & 0 \\
    \gamma_{14} & \gamma_{24} & 0 & 1 \\
  \end{array}\right)$ & $q^{1},q^{2},q^{3},q^{5}$ \\
  \hline
  $A_{3,6}\oplus A_{1}$ & $\left(\begin{array}{cccc}
    1 & 0 & 0 & \gamma_{14} \\
    0 & \gamma_{22} & 0 & \gamma_{24}\\
    0 & 0 & \gamma_{33} & 0 \\
    \gamma_{14} & \gamma_{24} & 0 & 1 \\
  \end{array}\right)$ & $q^{1},q^{2},q^{3},q^{5}$ \\
  \hline
  $A^{\alpha}_{3,7}\oplus A_{1}$ & $\left(\begin{array}{cccc}
    1 & 0 & 0 & \gamma_{14} \\
    0 & \gamma_{22} & 0 & \gamma_{24}\\
    0 & 0 & \gamma_{33} & 0 \\
    \gamma_{14} & \gamma_{24} & 0 & 1 \\
  \end{array}\right)$ & $q^{1},q^{3},q^{5},q^{6}$\\
  \hline
  $A_{3,8}\oplus A_{1}$ and $A_{3,9}\oplus A_{1}$ & $\left(\begin{array}{cccc}
    \gamma_{11} & 0 & 0 & \gamma_{14} \\
    0 & \gamma_{22} & 0 & \gamma_{24}\\
    0 & 0 & \gamma_{33} & \gamma_{34} \\
    \gamma_{14} & \gamma_{24} & \gamma_{34} & 1 \\
  \end{array}\right)$ & $q^{1},q^{2},q^{3},q^{4},q^{5},q^{6}$ \\
  \hline
  $A_{4,1}$ & $\left(\begin{array}{cccc}
    \gamma_{11} & \gamma_{12} & 0 & 0 \\
    \gamma_{12} & \gamma_{22} & 0 & 0\\
    0 & 0 & 1 & 0 \\
    0 & 0 & 0 & 1 \\
  \end{array}\right)$ & $q^{1},q^{3},q^{5}$ \\
  \hline
  $A^{\alpha}_{4,2}$ & $\left(\begin{array}{cccc}
    1 & \gamma_{12} & \gamma_{13} & 0 \\
    \gamma_{12} & 1 & \gamma_{23} & 0\\
    \gamma_{13} & \gamma_{23} & \gamma_{33} & \gamma_{34} \\
    0 & 0 & \gamma_{34} & \gamma_{44} \\
  \end{array}\right)$ & $q^{1},q^{2},q^{3},q^{4},q^{5},q^{6}$ \\
  \hline
\hline
\end{tabular}
\end{center}
\endlandscape
\newpage
\landscape
\begin{center}
\begin{tabular}{|c|c|c|}
\multicolumn{3}{c}{TABLE 2 (Continued)}\\
  \hline
  \hline
  Lie Algebra & (Possible) Reducible Form of $\gamma_{\alpha\beta}$ & Functionally Independent Invariants \\
  \hline
  \hline
  $A^{1}_{4,2}$ & $\left(\begin{array}{cccc}
    1 & 0 & \gamma_{13} & 0 \\
    0 & 1 & \gamma_{23} & 0\\
    \gamma_{13} & \gamma_{23} & \gamma_{33} & \gamma_{34} \\
    0 & 0 & \gamma_{34} & \gamma_{44} \\
  \end{array}\right)$ & $q^{1},q^{2},q^{3},q^{4},q^{5}$ \\
  \hline
  $A_{4,3}$ & $\left(\begin{array}{cccc}
    1 & \gamma_{12} & \gamma_{13} & 0 \\
    \gamma_{12} & 1 & 0 & 0\\
    \gamma_{13} & 0 & \gamma_{33} & 0 \\
    0 & 0 & 0 & \gamma_{44} \\
  \end{array}\right)$ & $q^{1},q^{2},q^{3},q^{5}$ \\
  \hline
  $A_{4,4}$ & $\left(\begin{array}{cccc}
    1 & 0 & 0 & 0 \\
    0 & \gamma_{22} & \gamma_{23} & 0\\
    0 & \gamma_{23} & \gamma_{33} & 0 \\
    0 & 0 & 0 & \gamma_{44} \\
  \end{array}\right)$ & $q^{1},q^{2},q^{3},q^{5}$ \\
  \hline
  $A^{\alpha,\beta}_{4,5}$ & $\left(\begin{array}{cccc}
    1 & \gamma_{12} & \gamma_{13} & 0 \\
    \gamma_{12} & 1 & \gamma_{23} & 0\\
    \gamma_{13} & \gamma_{23} & 1 & 0 \\
    0 & 0 & 0 & \gamma_{44} \\
  \end{array}\right)$ & $q^{1},q^{2},q^{3},q^{5}$ \\
  \hline
  $A^{\alpha,\alpha}_{4,5}$ & $\left(\begin{array}{cccc}
    1 & \gamma_{12} & 0 & 0 \\
    \gamma_{12} & 1 & 0 & 0\\
    0 & 0 & 1 & 0 \\
    0 & 0 & 0 & \gamma_{44} \\
  \end{array}\right)$ & $q^{1},q^{2}$ \\
  \hline
  $A^{\alpha,1}_{4,5}$ & $\left(\begin{array}{cccc}
    1 & 0 & 0 & 0 \\
    0 & 1 & \gamma_{23} & 0\\
    0 & \gamma_{23} & 1 & 0 \\
    0 & 0 & 0 & \gamma_{44} \\
  \end{array}\right)$ & $q^{1},q^{2}$ \\
  \hline
  $A^{1,1}_{4,5}$ & $\left(\begin{array}{cccc}
    1 & 0 & 0 & 0 \\
    0 & 1 & 0 & 0\\
    0 & 0 & 1 & 0 \\
    0 & 0 & 0 & \gamma_{44} \\
  \end{array}\right)$ & $q^{1}$ \\
  \hline
\hline
\end{tabular}
\end{center}
\endlandscape
\newpage
\landscape
\begin{center}
\begin{tabular}{|c|c|c|}
\multicolumn{3}{c}{TABLE 2 (Continued)}\\
  \hline
  \hline
  Lie Algebra & (Possible) Reducible Form of $\gamma_{\alpha\beta}$ & Functionally Independent Invariants \\
  \hline
  \hline
   $A^{\alpha,\beta}_{4,6}$ & $\left(\begin{array}{cccc}
    1 & \gamma_{12} & \gamma_{13} & 0 \\
    \gamma_{12} & 1 & 0 & 0\\
    \gamma_{13} & 0 & \gamma_{33} & 0 \\
    0 & 0 & 0 & \gamma_{44} \\
  \end{array}\right)$ & $q^{1},q^{3},q^{4},q^{5}$ \\
  \hline
  $A_{4,7}$ & $\left(\begin{array}{cccc}
    1 & 0 & 0 & 0 \\
    0 & \gamma_{22} & 0 & \gamma_{24}\\
    0 & 0 & \gamma_{33} & \gamma_{34} \\
    0 & \gamma_{24} & \gamma_{34} & \gamma_{44} \\
  \end{array}\right)$ & $q^{1},q^{2},q^{3},q^{4},q^{5}$ \\
  \hline
  $A_{4,8}$ & $\left(\begin{array}{cccc}
    \gamma_{11} & 0 & 0 & 0 \\
    0 & 1 & \gamma_{23} & \gamma_{24}\\
    0 & \gamma_{23} & 1 & \gamma_{34} \\
    0 & \gamma_{24} & \gamma_{34} & \gamma_{44} \\
  \end{array}\right)$ & $q^{1},q^{2},q^{3},q^{4},q^{5}$ \\
  \hline
  $A^{\beta}_{4,9}$ & $\left(\begin{array}{cccc}
    1 & 0 & 0 & 0 \\
    0 & 1 & \gamma_{23} & \gamma_{24}\\
    0 & \gamma_{23} & \gamma_{33} & \gamma_{34} \\
    0 & \gamma_{24} & \gamma_{34} & \gamma_{44} \\
  \end{array}\right)$ & $q^{1},q^{2},q^{3},q^{4},q^{5}$ \\
  \hline
  $A^{1}_{4,9}$ & $\left(\begin{array}{cccc}
    \gamma_{11} & 0 & 0 & 0 \\
    0 & 1 & 0 & 0\\
    0 & 0 & 1 & \gamma_{34} \\
    0 & 0 & \gamma_{34} & \gamma_{44} \\
  \end{array}\right)$ & $q^{1},q^{2},q^{3}$ \\
  \hline
  $A^{0}_{4,9}$ & $\left(\begin{array}{cccc}
    1 & 0 & 0 & 0 \\
    0 & 1 & \gamma_{23} & \gamma_{24}\\
    0 & \gamma_{23} & \gamma_{33} & \gamma_{34} \\
    0 & \gamma_{24} & \gamma_{34} & \gamma_{44} \\
  \end{array}\right)$ & $q^{1},q^{2},q^{3},q^{4},q^{5}$ \\
  \hline
  $A_{4,10}$ & $\left(\begin{array}{cccc}
    \gamma_{11} & 0 & 0 & 0 \\
    0 & 1 & 0 & \gamma_{24}\\
    0 & 0 & \gamma_{33} & \gamma_{34} \\
    0 & \gamma_{24} & \gamma_{34} & \gamma_{44} \\
  \end{array}\right)$ & $q^{1},q^{2},q^{3},q^{4},q^{5}$ \\
  \hline
\hline
\end{tabular}
\end{center}
\endlandscape
\newpage
\landscape
\begin{center}
\begin{tabular}{|c|c|c|}
\multicolumn{3}{c}{TABLE 2 (Continued)}\\
  \hline
  \hline
  Lie Algebra & (Possible) Reducible Form of $\gamma_{\alpha\beta}$ & Functionally Independent Invariants \\
  \hline
  \hline
  $A^{\alpha}_{4,11}$ & $\left(\begin{array}{cccc}
    \gamma_{11} & 0 & 0 & 0 \\
    0 & 1 & 0 & \gamma_{24}\\
    0 & 0 & \gamma_{33} & \gamma_{34} \\
    0 & \gamma_{24} & \gamma_{34} & \gamma_{44} \\
  \end{array}\right)$ & $q^{1},q^{3},q^{4},q^{5},q^{6}$ \\
  \hline
  $A_{4,12}$ & $\left(\begin{array}{cccc}
    1 & 0 & 0 & 0 \\
    0 & \gamma_{22} & \gamma_{23} & \gamma_{33}\\
    0 & \gamma_{23} & \gamma_{33} & \gamma_{34} \\
    0 & \gamma_{24} & \gamma_{34} & \gamma_{44} \\
  \end{array}\right)$ & $q^{1},q^{2},q^{3},q^{4},q^{5},q^{6}$ \\
  \hline
\hline
\end{tabular}
\end{center}
\endlandscape
\vspace*{1.5cm} \noindent \textbf{\large{Acknowledgements}}\\
One of us (G.O. Papadopoulos) is currently a scholar of the Greek
State Scholarships Foundation (I.K.Y.) and acknowledges the
relevant financial support.

\end{document}